\documentclass[english,aps,prper,reprint,superscriptaddress,floatfix]{revtex4-2}

\usepackage{graphicx}
\usepackage[colorlinks=true,allcolors=blue]{hyperref}
\usepackage{multirow}
\usepackage{rotating}
\usepackage{rotate}
\usepackage{lineno,xcolor}
\usepackage{amsmath,amssymb}
\usepackage{array}
\usepackage{comment}

\makeatletter
\newcommand*\bigcdot{\mathpalette\bigcdot@{.5}}
\newcommand*\bigcdot@[2]{\mathbin{\vcenter{\hbox{\scalebox{#2}{$\m@th#1\bullet$}}}}}
\makeatother

\begin{document}
\title{Replicating analyses of item response curves using data from the Force and Motion Conceptual Evaluation}

\author{Connor J.\ Richardson}
\affiliation{Department of Physics and Astronomy, Rowan University, Glassboro, NJ 08028, USA}

\author{Trevor I.\ Smith}
\affiliation{Department of Physics and Astronomy, Rowan University, Glassboro, NJ 08028, USA}
\affiliation{Department of STEAM Education, Rowan University, Glassboro, NJ 08028, USA}

\author{Paul J.\ Walter}
\affiliation{Department of Mathematics, St.~Edward's University, Austin, TX 78704, USA}

\begin{abstract}
Ishimoto, Davenport, and Wittmann have previously reported analyses of data from student responses to the Force and Motion Conceptual Evaluation (FMCE), in which they used item response curves (IRCs) to make claims about American and Japanese students' relative likelihood to choose certain incorrect responses to some questions. We have used an independent data set of over 6,500 American students' responses to the FMCE to generate IRCs to test their claims. Converting the IRCs to vectors, we used dot product analysis to compare each response item quantitatively. For most questions, our analyses are consistent with Ishimoto, Davenport, and Wittmann, with some results suggesting more minor differences between American and Japanese students than previously reported. 
We also highlight the pedagogical advantages of using IRCs to determine the differences in response patterns for different populations to better understand student thinking prior to instruction.  
\end{abstract}

\maketitle

\section{Introduction}
Research-based multiple-choice assessment instruments, such as the Force Concept Inventory (FCI \cite{Hestenes1992}) and Force and Motion Conceptual Evaluation (FMCE \cite{Thornton1998}), are ubiquitous in physics education research. These tools have been used as standard measures for research, instruction, and programmatic assessment purposes for the past three decades \cite{Madsen2017, Madsen2019}. Conceptual assessments have been beneficial for illustrating the need for and benefit of research-based instructional materials and strategies \cite{Hake1998, Thornton2009,Smith2014, Engelhardt2004,Ding2006}.

Although the FCI and FMCE have been in use for over 20 years, research persists on how to use these assessments and interpret the results. 
Ishimoto, Davenport, and Wittmann compared preinstruction FMCE student responses from a population of American introductory physics students to preinstruction responses from a population of Japanese introductory physics students using both overall score distributions and item response curve (IRC) analyses \cite{Ishimoto2017}.  IRCs show the fraction of students who selected each answer choice to a particular item as a function of their total score on the test \cite{Morris2006,Morris2012,Walter2016,Morris2017}. These analyses can show if different groups of students (e.g., high-scoring or low-scoring) select different incorrect answer choices and help reveal trends evident in students' patterns of correct and incorrect answer choices. 

Our goal is to replicate the work of Ishimoto, Davenport, and Wittmann from Ref.\ \cite{Ishimoto2017}, which we will refer to as IDW. Our data are comprised of preinstruction responses to the FMCE from 6,584 American introductory physics students from a variety of colleges and universities. We carefully compare the score distribution from our data to those published by IDW, and we use a recently developed method for comparing IRCs quantitatively between the three populations \cite{Walter2021}. Our primary goal is to determine how well our results agree with their results from American students, which will indicate whether their results are more generally representative of American students. We also compare our results with their Japanese students' results to determine if we see the same discrepancies that they report between American and Japanese populations. 

IDW found that the preinstruction score distributions for American and Japanese students are fairly similar in shape, but that the scores for Japanese students are statistically significantly higher than those of American students ($11.22 \pm 0.19$ vs.\ $9.05 \pm 0.15$, $t=8.8$, $p<0.005$). Using qualitative visual comparisons, IDW concluded that the patterns of correct and incorrect answer choices shown on IRCs did not differ greatly between the American and Japanese student populations for most FMCE items. They reported some isolated cases of discrepancies in the answer patterns shown in the IRCs, which they attribute to cultural differences between the two populations, e.g., children growing up in Japan do not have as much experience with flipping coins as children in the USA, and American students typically have more experience driving and riding in cars than Japanese students.

Based on the findings of IDW, and our overall goals for replicating their work, we answer the following research questions:
\begin{enumerate}
    \item Does the distribution of preinstruction FMCE scores from our data set match those from the IDW American data set and support their claim that Japanese students have higher preinstruction scores than American students?
    \item Do the IRCs for our data set match those from the IDW American data set, indicating that students chose the same responses at similar rates?
    \item Are the IRCs for our data set different from those from the IDW Japanese data set in the same ways they previously reported for American and Japanese students? 
    \item What additional information about the similarities or differences between IRCs can we reveal by using rigorous quantitative comparisons?
\end{enumerate}
Without access to independent data from Japanese students, we cannot fully replicate the IDW study. However, we feel confident that using an independent data set of American students that is more than twice as large as in their original study allows us to make strong claims about the veracity of their results from American students, which will then, in turn, allow us to evaluate the strength of their overall claims.

\section{Item Response Curves}
Morris et al.\ introduced IRCs as a simplified form of item response theory (IRT) that uses the total test score as an independent variable, rather than the IRT latent trait of ability level, which, by definition, is unmeasurable \cite{Morris2006, Morris2012}. IRC analyses are similar to the IRT nominal response model but require far less computational power \cite{Bock1972, Suh2010, Smith2020rank, Eaton2019irt}.

\subsection{Reading an IRC}
\label{readingirc}
\begin{figure}[htp!]
  \includegraphics[width=\linewidth]{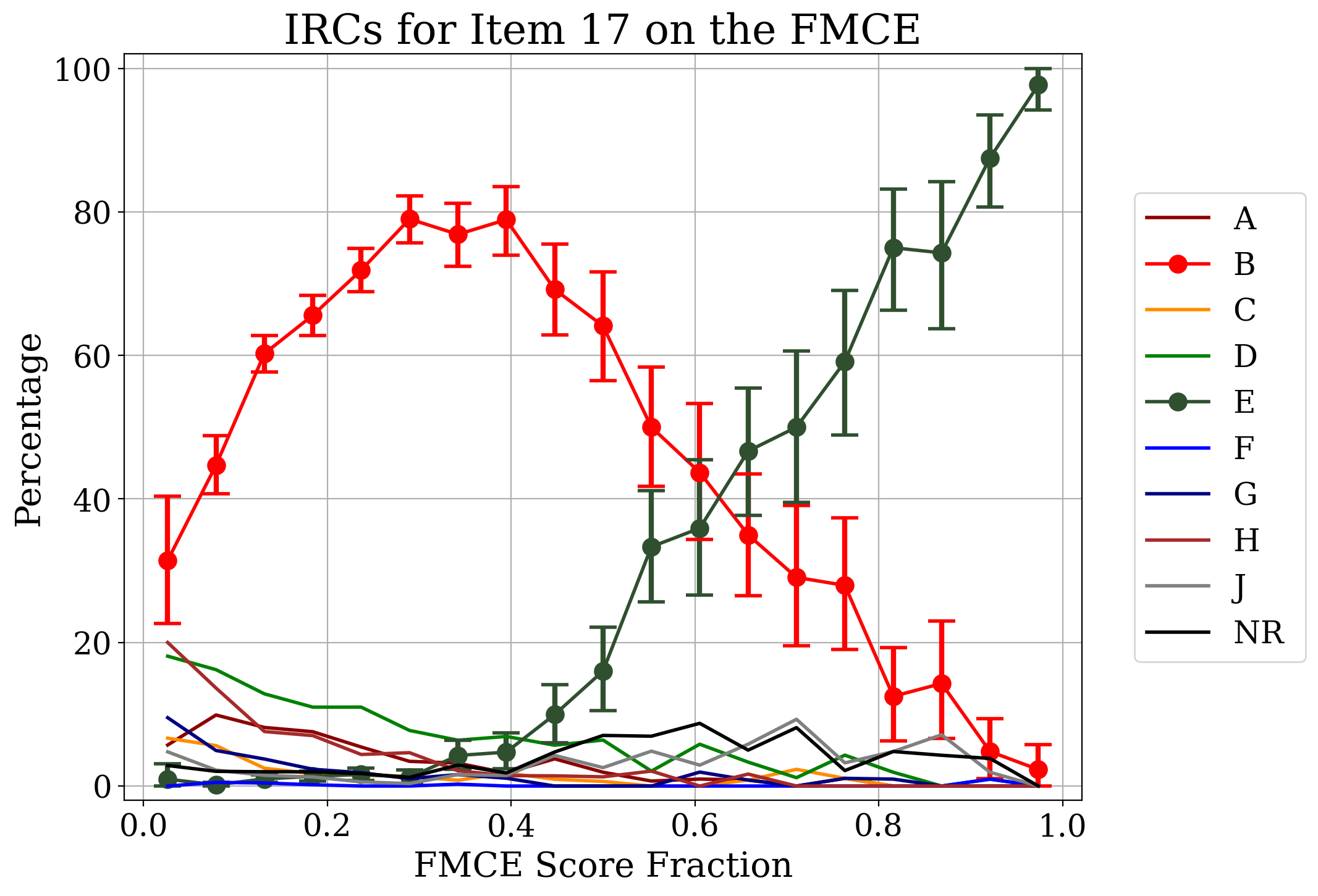}
  \caption{Item response curves for 6,584 American students (the RSW data set) for item 17 on the FMCE. NR = no response. The error bars on E (the correct answer) and B (the most commonly chosen incorrect answer) are the 95\% confidence intervals of 10,000 bootstrapped sample data sets based on the RSW data. For visual clarity, specific plot markers and error bars are omitted for less frequently chosen answers.}
  \label{fig:ircs_q17_rsw}
\end{figure}

To illustrate the information shown in IRCs, Fig.~\ref{fig:ircs_q17_rsw} shows an example of students' response patterns for Item 17 on the FMCE.  There is one curve for each answer choice, which shows the percentage of students selecting a particular answer choice as a function of respondents' fractional overall score. The line segments connecting the data points only serve as a visual aid.  The correct answer choice is E (shown in dark green); the correct answer on any IRC plot may be seen as having low frequency with low-scoring students and high frequency with high-scoring students. Answer choice B (shown in red) is the most common incorrect answer choice (shown by higher frequencies than all other incorrect answers), and it is consistent with the common idea that the total force on an object is proportional to the object's velocity. Answer choice B also has an intermediate maximum, indicating that students are most likely to choose B when they have a moderate level of understanding (indicated by fractional score) \footnote{P\'{e}rez-Lemonche et al.\ have suggested that an intermediate maximum may indicate that students with a particular misconception are attracted to that answer choice \cite{Perez2019}.}. 

An important caveat when reading IRCs for preinstruction FMCE responses is that the score distribution is not uniform. As shown in Fig.\ \ref{fig:score_distributions}, the mode of the distribution occurs at a fractional score of $\sim 0.15-0.2$, and a large majority of students earn scores $< 0.4$. In this way, IRCs cannot directly show students' overall likelihood of choosing each answer choice. 
Additionally, answer choice J is the equivalent of ``none of the above'' for all items on the FMCE; therefore, we do not consider students who choose J to be a coherent group: we know that they think that none of the other answers is correct, be we cannot claim that they all agree on what a correct answer would be. 

\subsection{IRC vs.\ IRT}
\label{ircvirt}
Item response theory (IRT) has a solid theoretical foundation that has been developed over the past several decades based on the notion that student responses to multiple-choice tests are dependent on a latent trait that cannot be measured directly (in our context, this would be an overall understanding of introductory mechanics) \cite{deAyala2008,Bock1972,Thissen2010,Bock2007,Suh2010}. In contrast, item response curves have been in use for about 15 years, and they are no more complicated than reporting population-specific conditional probabilities of students choosing a particular response given their score on the test; students' overall scores substitute for the IRT latent trait of ability level \cite{Morris2006}. As such, IRCs can be viewed as a simplified form of IRT with less statistical power. 

Using a rigorous IRT-based analysis would provide more statistical power for any of our claims, and recent work has shown how the nominal response model can be applied to FMCE data to determine a ranking for all incorrect responses to each question \cite{Smith2020rank}; however, there are benefits to using IRC-based analyses instead. IRT analyses require data sets at least ten times as large as the number of estimated parameters to have robust statistical power \cite{deAyala2008}: a minimum of 3,000 students for the FCI and 7,220 for the FMCE. Additionally, performing IRT analyses requires fairly sophisticated computational methods. Software exists to perform these calculations (e.g.\ the mirt package for R \cite{r, Chalmers2012}), but IRC analyses are much simpler and require only basic spreadsheet operations to perform. As such, IRC analyses are more accessible for researchers with smaller data sets and those who prefer more straightforward computational methods. We have chosen to use IRC analyses for this work because our goal is to replicate the IDW study \cite{Ishimoto2017} as closely as possible.

\section{Methods}
\label{methods}
Many of our methodological choices were taken directly from the IDW study \cite{Ishimoto2017}: cleaning the data set based on rules for removing some incomplete response sets, calculating an overall FMCE score for each student based on the common 37-point scoring scheme, creating two-score bins for generating IRCs, and using statistical methods to compare overall score distributions and qualitative visual comparisons of IRCs. We augmented these methods by performing additional statistical analyses to compare score distributions and utilizing quantitative analyses to compare IRCs. 

Our raw data consist of preinstruction item responses to the FMCE gathered from 7,325 students at various colleges and universities in the United States known to be disjoint from IDW's data pool. One of our major sources is PhysPort's Data Explorer, a large online database of anonymous data uploaded by instructors \cite{physportde}. We know little more than that these students were enrolled in an introductory physics course in the USA, and that they were given the FMCE before and after instruction. We do not know their majors, previous levels of physics taken, or demographic information; IDW report having some of this information \cite{Ishimoto2017}. Approximately 1,000 students in our data set come from four known data sources: two public state universities, a two-year college, and a selective private liberal arts college. Other analyses of some of these data sources have previously been reported \cite{Smith2014,Griffin2016,Smith2017}. Some information is known about the instructional methods used at these four institutions, but no demographic information is available about the students. 

\subsection{Cleaning and Analyzing the Data Set}
We mirrored IDW's methods for cleaning the data. As such, we only included response sets that included answers to at least one item in each of the previously defined item clusters (Force Sled, Reversing Direction, Force Graphs, Acceleration Graphs, Velocity Graphs, Newton III, and Energy \cite{Smith2008, Ishimoto2017}). We also omitted response sets that had more than 6 unanswered items in total. IDW mentions having the same student appear twice and choosing to use their first responses. We did not have a method to test if a student appeared twice in our data. After cleaning, we removed the responses of 741 students leaving 6,584 to be analyzed. We refer to this as the RSW data set.

To compare students, we generated a score in the same manner as IDW, who followed the recommended scoring procedure of Thornton and Sokoloff \cite{Thornton1998,Thornton2009}: items 5, 6, 15, 33, 35, 37, and 39 are not scored; 
items 8--10, 11--13, and 27--29 are each treated as sets in which students earn two points for answering all three items in a set correctly, and zero points if any item is answered incorrectly. Thus, we used the same 40 items to calculate a student's score, with a maximum score of 37.   

We used a two-score binning method to create our IRCs, so students who scored a 0 or 1, 2 or 3, etc., were plotted together. Binning in this way reduces the standard error of the plots because it increases the sample size per point \cite{Ishimoto2017}. We calculated the fraction of students who chose each answer choice for each score bin and plotted these as IRCs. To estimate the uncertainty in our (RSW) IRCs, we used 10,000 bootstrapped samples to generate a distribution of values for each data point \cite{boot2020, boot1997}. We include error bars on our IRCs to indicate the central 95\% of each distribution \footnote{For visual clarity, we only include error bars on curves with values above 20\%. See Fig.\ \ref{fig:ircs_q17_rsw} for an example.}.

\subsection{Comparing Data Sets}
\label{sec:compare}

\subsubsection{Comparing Score Distributions}
\label{sec:scores}
Ishimoto, Davenport, and Wittmann reported the score distributions for both of their data sets (as a bar graph in Fig.\ 4 of IDW \cite{Ishimoto2017}), as well as the population size, mean score, median score, and standard deviation for each (in Table II of IDW \cite{Ishimoto2017}). Using the image processing software ImageJ (version 1.53) \cite{Schneider2012}, we analyzed the IDW score distribution bar graph to estimate the fraction of students in each data set that earned each score on the FMCE. Combing this information with the reported descriptive statistics allowed us to determine the number of students that earned each score and recreate the score distributions from IDW \cite{Ishimoto2017}. 

We calculated the same descriptive statistics for our data that were reported in IDW, but due to the non-parametric nature of the score distributions, we chose comparison methods that deviate somewhat from theirs. We compared our score distribution to both of the IDW distributions using a Kruskal-Wallis rank-sum test, and we performed post hoc pairwise comparisons with the Wilcoxon rank-sum test, using the BH p-value correction method. Additionally, in recognizing that p-values alone are not sufficient for determining whether or not a real effect exists \cite{Wasserstein2019,Wasserstein2016}, we chose to compute the effect sizes of the differences between the data sets using Hedges' $g$ in order to get a more complete picture of any differences. Analyses used the R statistical software platform \cite{r}.

\subsubsection{Treating IRCs as Vectors and Taking Dot Products}
\label{sec:dotproduct}
Walter, Nuhfer, and Suarez provide a means of quantitatively comparing the IRCs of two populations by treating the information in IRCs as multidimensional vectors and computing a dot product as a metric of similarity \cite{Walter2021}.  They compared the IRCs of groups with differing demographics (e.g., gender, ethnicity) using data collected from the 25-item Science Literacy Concept Inventory \cite{Nuhfer2016}. For each item in our work, a population has 19 associated IRC vectors, one for each score bin; each vector contains one component for each answer choice and is normalized over all answer choices for that item (5--9 for the FMCE). The IRC dot product is the weighted (based on the number of respondents per score bin) average of dot products of the normalized IRC vectors of two populations. Since all IRC vector components are $\ge 0$, the values of the IRC dot product range from 0 (the IRCs are completely different) to 1 (the IRCs are identical) \footnote{The actual minimum value of an IRC dot product is significantly higher than 0, which we discuss below}.

To compare two populations using IRC dot products, one must know each data point's precise value for each IRC. One author of the IDW study was gracious enough to share their results with us to make these comparisons.

In this work, we also compute the IRC dot products between each of the 10,000 bootstrapped samples based on the RSW data set and the American and Japanese data sets from IDW \cite{Ishimoto2017}, providing a distribution of values of the IRC dot product for each item. We used these distributions to create error bars representing the central 95\% of each distribution, which we will refer to as the ``IRC dot product confidence interval.''

\subsubsection{Randomized Trial Confidence Interval}
\label{sec:random_interval}

To have a means of determining whether a value of an IRC dot product is likely to arise from random chance, Walter, Nuhfer, and Suarez introduce what we will refer to as the ``randomized trial confidence interval'' \cite{Walter2021}.  To construct this interval, they determine the range of IRC dot product values from 10,000 randomized trials of simulated students. 

Each randomized trial involves creating two simulated populations, each with the same number of students as their respective corresponding real populations ($a$ and $b$). Each simulated student is assigned to a scoring bin using weighted random sampling based on the corresponding real population's score distribution.  This process ensures each simulated population's score distribution will closely mimic the corresponding real students' score distribution.  Each simulated student is then assigned an answer choice using weighted random sampling of the overall population's probabilities ($a$ and $b$ combined) for students in the same score bin. The simulated populations thus select answer choices using the same probability distribution. 

Each randomized trial is completed by taking the simulated populations' IRC dot product.  After completing 10,000 randomized trials, a range of outcomes is constructed, capturing 95\% of the randomized trials (from the 2.5 to 97.5 percentiles).  
For any given item, when the real population's IRC dot product falls within the 95\% randomized trial confidence interval, then the differences in IRCs for the real populations may result from random chance.

\subsubsection{Comparing to Purely Random Data}
\label{sec:pure_random}
Given that the correct answer IRC for any population must have low frequencies at low scores and high frequencies for high scores, and given that all IRCs for incorrect answers must have low frequencies at high scores, the IRC dot product between any two populations will never be zero. To establish a reasonable lower bound for dot-product comparisons, we created three simulated student populations equal in number to RSW, IDWA, and IDWJ, respectively. Each of these simulated data sets was created under the assumption that respondents answered every item randomly. By taking the IRC dot product of a simulated student population with its associated real student population, we can estimate a minimum baseline value of the IRC dot product for completely unrelated data sets.  The item and population-specific baseline dot product value can depend on the population's size, the number of answer choices, the number of answer choices serving as effective distractors, and the item difficulty.

\section{Score distributions are not different across populations}
\label{sec:scoreResults}
\begin{figure}[tb]
  \includegraphics[width=\linewidth]{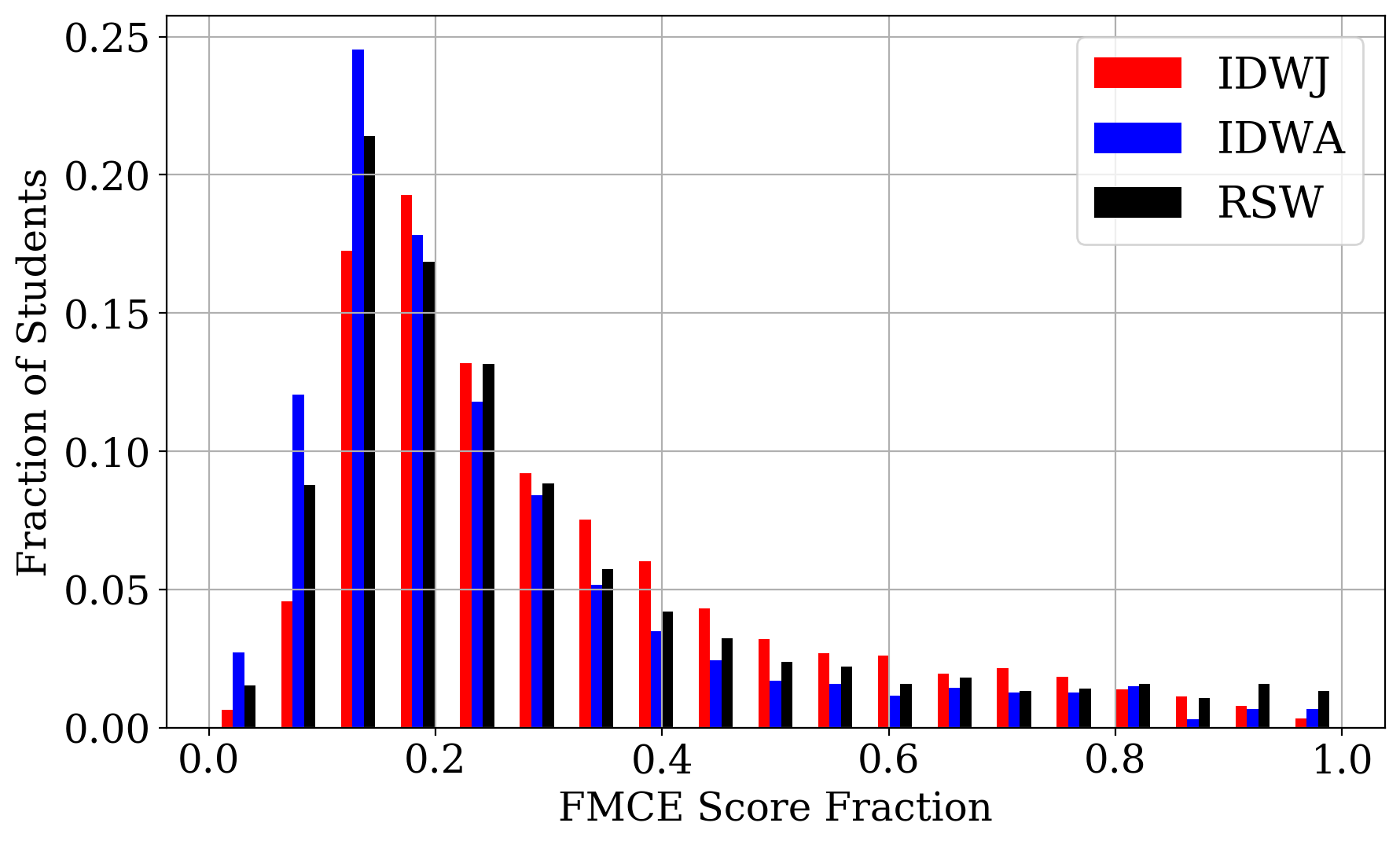}
  \caption{The fractional score distributions on the FMCE are shown for three separate datasets: 1,531 Japanese students (IDWJ) and 2,347 American students (IDWA) used in Ishimoto, Davenport, and Wittmann \cite{Ishimoto2017}; 6,584 American students (RSW) obtained from PhysPort and other sources \cite{physportde,Smith2014,Griffin2016,Smith2017}.}
  \label{fig:score_distributions}
\end{figure}

Ishimoto, Davenport, and Wittmann report that their data sets' score distributions are statistically different according to a two-tailed $t$-test ($t=8.8$, $p<0.005$) \cite{Ishimoto2017}. Figure \ref{fig:score_distributions} shows the FMCE preinstruction score distributions for all three data sets: our data (labeled RSW), the IDW American data set (IDWA), and the IDW Japanese data set (IDWJ). Table \ref{tab:scores} shows the descriptive statistics for these distributions.

\begin{table}[bt]
    \centering
    \caption{Descriptive statistics for all three data dets: our data (RSW), the American data from IDW (IDWA), and the Japanese data from IDW (IDWJ).}
    \begin{ruledtabular}
    \begin{tabular}{lccc}
         Statistic & RSW & IDWA & IDWJ \\
         \hline
         Sample size & 6,584 & 2,348 & 1,531 \\
         Mean score & 10.46 & 9.05 & 11.22\\
         Median score & 8 & 7 & 9 \\
         Standard deviation & 8.20 & 7.36 & 7.56\\
    \end{tabular}
    \end{ruledtabular}
    \label{tab:scores}
\end{table}

The results from a Kruskal-Wallis rank-sum test show a statistically significant main effect between the three data sets, $\chi^2(2) = 143.66$, $p < 0.001$; moreover, the pairwise Wilcoxon rank-sum tests indicate that each data set is significantly different from each other data set for all three pairs, with $p < 0.001$. In contrast, calculations of Hedges' $g$ show that the effect sizes of these differences are fairly small: $g = 0.18$ for RSW -- IDWA, $g = 0.09$ for IDWJ -- RSW, and $g = 0.29$ for IDWJ -- IDWA \footnote{Cohen provided rough definitions of effect sizes of 0.2 being small, 0.5 being moderate, and 0.8 being large \cite{Cohen1988}. Recent work suggests that average effect sizes for the FCI and the FMCE are around $g = 1.0$ when comparing pre- and post-test data sets \cite{Nissen2020}. Based on both of these frames of reference, we consider our results of $g < 0.3$ to be quite small.}. The results from the pairwise Wilcoxon rank-sum tests agree with those of IDW, but we disagree with the interpretation that American students' score distribution differs meaningfully from that of Japanese students. The fact that the mean and median scores of our data set fall in between those of the IDW data sets, and the fact that the effect sizes between each pair of distributions are small, suggest that the score distributions of all three data sets are not meaningfully different (despite the statistical significance reported by the Kruskal-Wallis and Wilcoxon rank-sum tests) \footnote{We believe the statistical significance of these differences may be the result of having very large data sets.  Small differences may be categorized as statistically significant, even though they may not be particularly meaningful in the context of the measured outcome  \cite{Wasserstein2016,Wasserstein2019}}. The similarities between score distributions may be seen in Fig.\ \ref{fig:score_distributions} with all three largely overlapping each other and showing similar shapes.

\section{Item Response Curve Results}
\label{sec:ircResults}

\subsection{Results reported by Ishimoto, Davenport, and Wittmann}
Based on qualitative comparisons of their IRCs, Ishimoto, Davenport, and Wittmann reported that most IRCs are quite similar for their American and Japanese data sets \cite{Ishimoto2017}; however, they reported several notable exceptions. 
\begin{itemize}
    \item On items 3 and 7 (a sled being pushed and slowing down), the most common incorrect response for both data sets is that the force is in the direction of motion and decreasing, but Japanese students are more likely than American students to choose an accompanying force that is opposite the direction of motion and increasing.
    \item Items 11, 12, 13 (force on a coin flipped vertically in the air) appeared more difficult for the Japanese students than the American students. In contrast, both groups did equally well on isomorphic items 8, 9, and 10, which use a toy car moving up and down a ramp.
    \item Items 16 and 18  ``appear to be slightly easier for the American students'' \cite{Ishimoto2017}. These items require students to choose a graph of force vs.\ time for a toy car moving to the right and either speeding up or slowing down, respectively.
    \item Item 22 (choosing a graph of acceleration vs.\ time for a toy car moving to the right and speeding up) was easier for American students than Japanese students.
    \item Items 30--32 and 34 (interaction forces during collisions) differed between American and Japanese students. American students selecting an incorrect answer are likely to use mass-dependence reasoning on item 30 (larger truck and smaller car moving at the same speed toward each other) and action-dependence reasoning on items 32 and 34 (a moving car hits a stationary truck of either greater or equal mass, respectively) \cite{Smith2008}. By contrast, Japanese students are much more likely to answer these questions correctly. One exception is that American students are similarly likely to answer item 31 correctly (a small fast car collides with a large slow truck); this may relate to a tendency of some students to see the opposite effects of mass-dependence and action-dependence as balancing to result in equal forces \cite{Smith2008}.
    \item Items 44--47 (selecting velocity or kinetic energy of a sled after going down a hill) are easier for Japanese students than American students.
\end{itemize}

To test these claims, we report the results from our dot product analyses for pairwise comparisons of the three data sets and the dot product analyses for each of the data sets compared to a data set of entirely random responses. For each pairwise comparison of data sets, we also show the range of IRC dot product values expected if the differences between the data sets resulted from random chance. We also examine the nature of the differences between data sets for the individual items listed in IDW as notably different across data sets.

\subsection{IRC Dot Product Results and Comparisons to Randomly Selected Responses}
\label{sec:dotResults}
\begin{figure}[tb]
  \includegraphics[width=\linewidth]{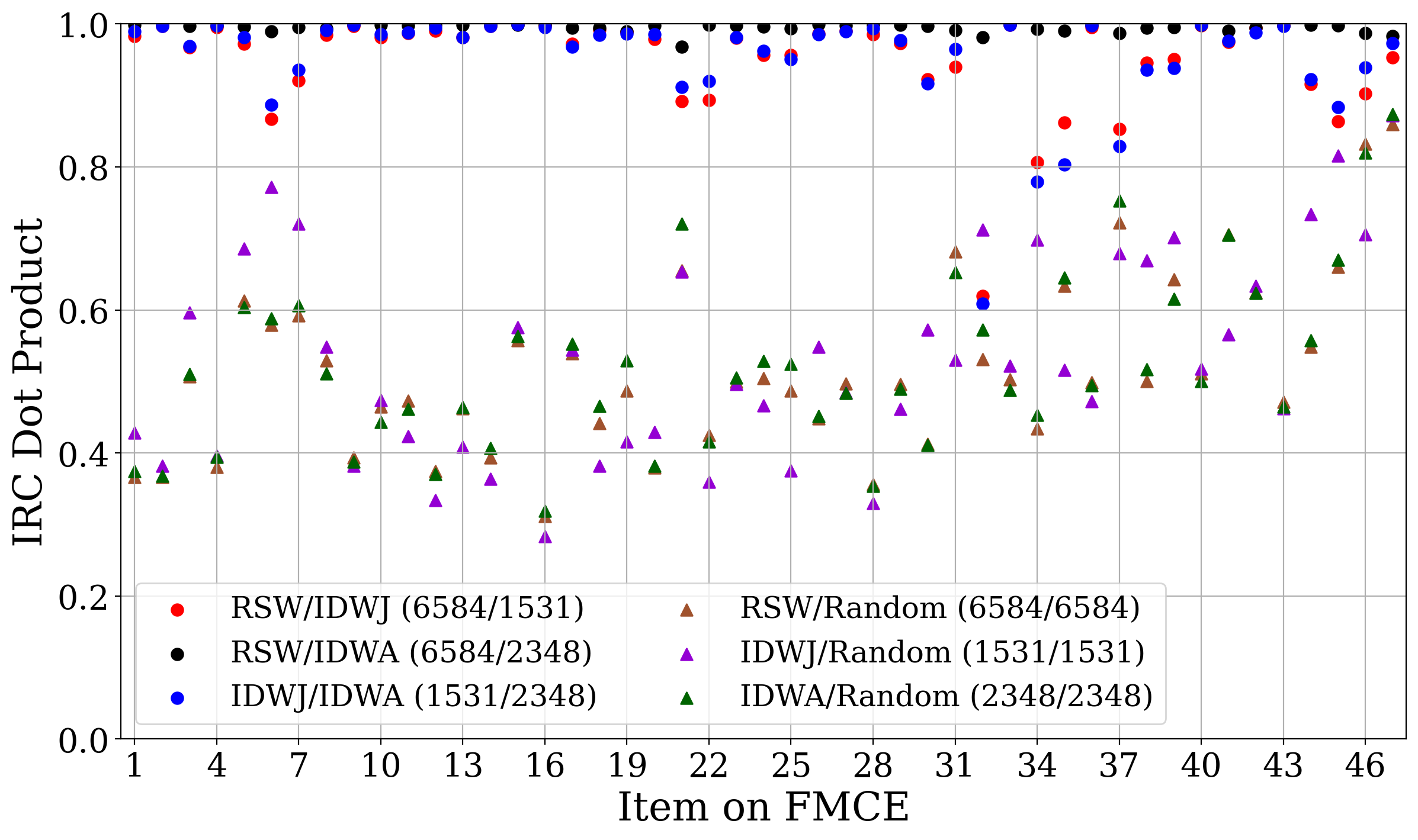}
  \caption{The IRC dot products for all three pairs of populations shown in Figures \ref{fig:dot_IDWJ_IDWA}, \ref{fig:dot_RSW_IDWJ}, and \ref{fig:dot_RSW_IDWA}.  The IRCs of the American students (RSW and IDWA) are quite similar, and thus the IRC dot product values of the Japanese students (IDWJ) with either set of American students (RSW or IDWA) are similar for each item. IRC dot products are also shown for each population compared to a simulated set of random responses.}
  \label{fig:dot_all}
\end{figure}

\begin{figure}[tb] 
  \includegraphics[width=\linewidth]{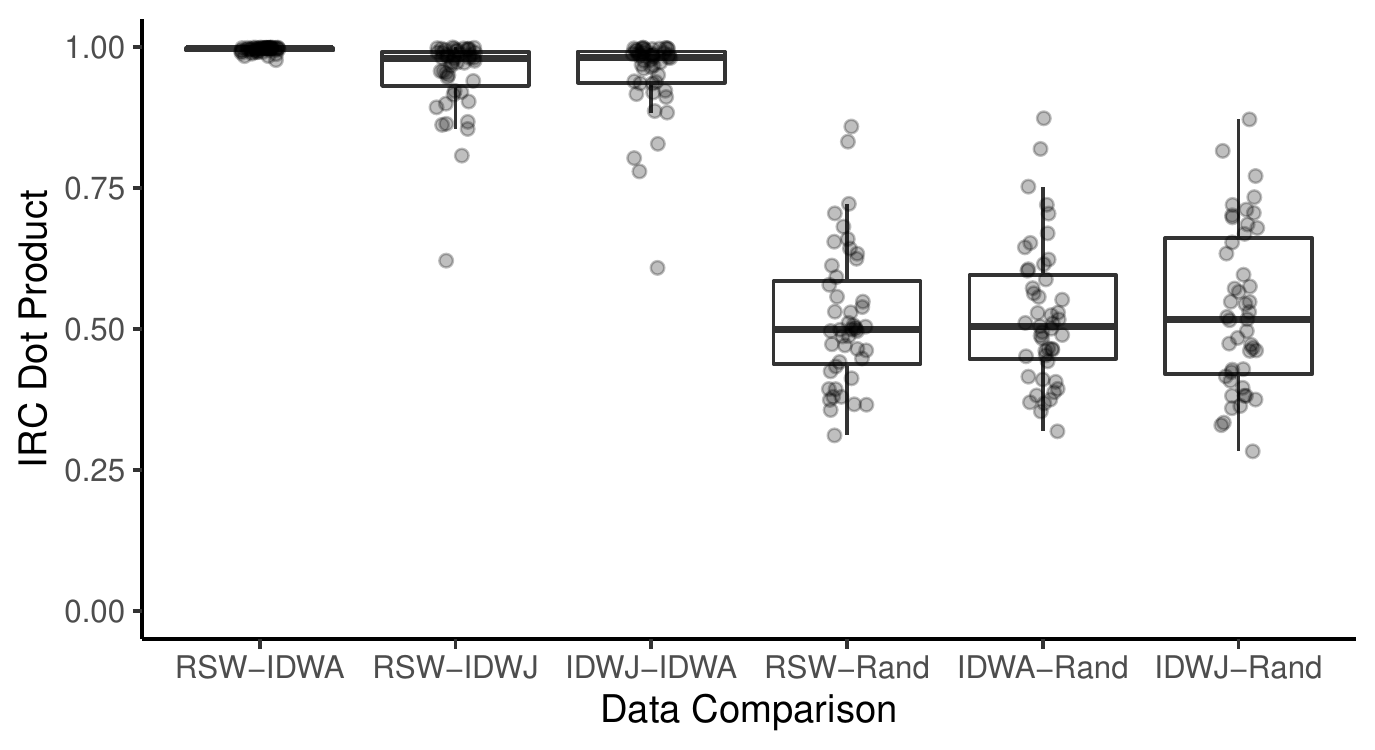}
  \caption{Distributions of the IRC dot product values for each of the comparisons shown in Fig.\ \ref{fig:dot_all}. The RSW-IDWA comparison has a very narrow distribution near unity. The RSW-IDWJ and IDWJ-IDWA distributions are very similar and lower than RSW-IDWA.  Each comparison to a simulated data set of random responses has a distribution that is noticeably lower than any of the comparisons between real data sets.}
  \label{fig:dot_all_box}
\end{figure}

Figure \ref{fig:dot_all} shows the IRC dot products for each item for each of the three pairs of populations: RSW-IDWJ; RSW-IDWA; and IDWJ-IDWA. Based on the similarities of the score distributions, one might expect that the RSW data set would produce IRCs that are equally similar to those of the IDWA and IDWJ data sets, but Fig.\ \ref{fig:dot_all} shows us that this is not the case. The IRC dot product values for each item for the RSW-IDWA comparison (black circles) show that the two American populations' IRCs are highly similar to each other, with all values within the range [0.9771, 0.9996].  Consequently, we find that values of the IRC dot product for each item for the RSW-IDWJ and the IDWJ-IDWA pairs are quite close to each other, both with much broader distributions of IRC dot product values: RSW-IDWJ values fall within the range [0.6214, 0.9997], and IDWJ-IDWA values are within the range [0.6087, 0.9993]. The boxplots in Fig.\ \ref{fig:dot_all_box} show the distribution of IRC dot product values for each comparison.

Figures \ref{fig:dot_all} and \ref{fig:dot_all_box} also show the results for each population (RSW, IDWA, and IDWJ) paired with an equal number of simulated students who are randomly guessing. For the vast majority of items, the comparisons within the three experimental data sets have considerably higher IRC dot product values than comparisons between any experimental data sets and the random data set; i.e., they are much closer to each other than with a random data set. One major exception to this trend is item 32, for which the IRC dot products between IDWJ and either RSW or IDWA are around 0.61, well within the range of IRC dot product values involving random data. Some other items, such as 45 and 46, have IRC dot product values between data sets that approach those involving random data, but they are still distinct.

The comparisons to simulated random data sets in Figs.\ \ref{fig:dot_all} and \ref{fig:dot_all_box} provide essential information and context when interpreting the meaning of the IRC dot product values. Theoretically, the value of a dot product must fall within the range [0, 1]; however, the lowest values seen in Fig.\ \ref{fig:dot_all} are slightly below 0.3 (item 16). Also, the IRC dot products' values may be affected by the structure of the test itself. Items 1--29 have eight or nine possible answer choices for each item, but items 44--47 have only five answer choices. As such, simulated random responses are more likely to match up with actual data for items 44--47, and we see larger values (in general) of the IRC dot products comparing random responses for those items (none is below 0.5). The number of answer choices, however, is not the only factor related to the IRC dot products with random data: items 1--4 have lower values than items 5--7, even though they all have eight possible answer choices and involve the same physical scenario (a sled being pushed across an icy lake).  The values of the IRC dot product of a real population with purely random data depend on the shape of the real populations' IRCs and the real populations' score distribution.  For example, purely random guessing will not mimic the IRC for an answer choice that acts as a distractor with an intermediate maximum.    

\subsection{IRC Dot Products Compared to Uncertainty Ranges}
\label{sec:dotRange}

\subsubsection{Reexamining the comparisons of Ishimoto, Davenport, and Wittmann}
\begin{figure}[tb]
  \includegraphics[width=\linewidth]{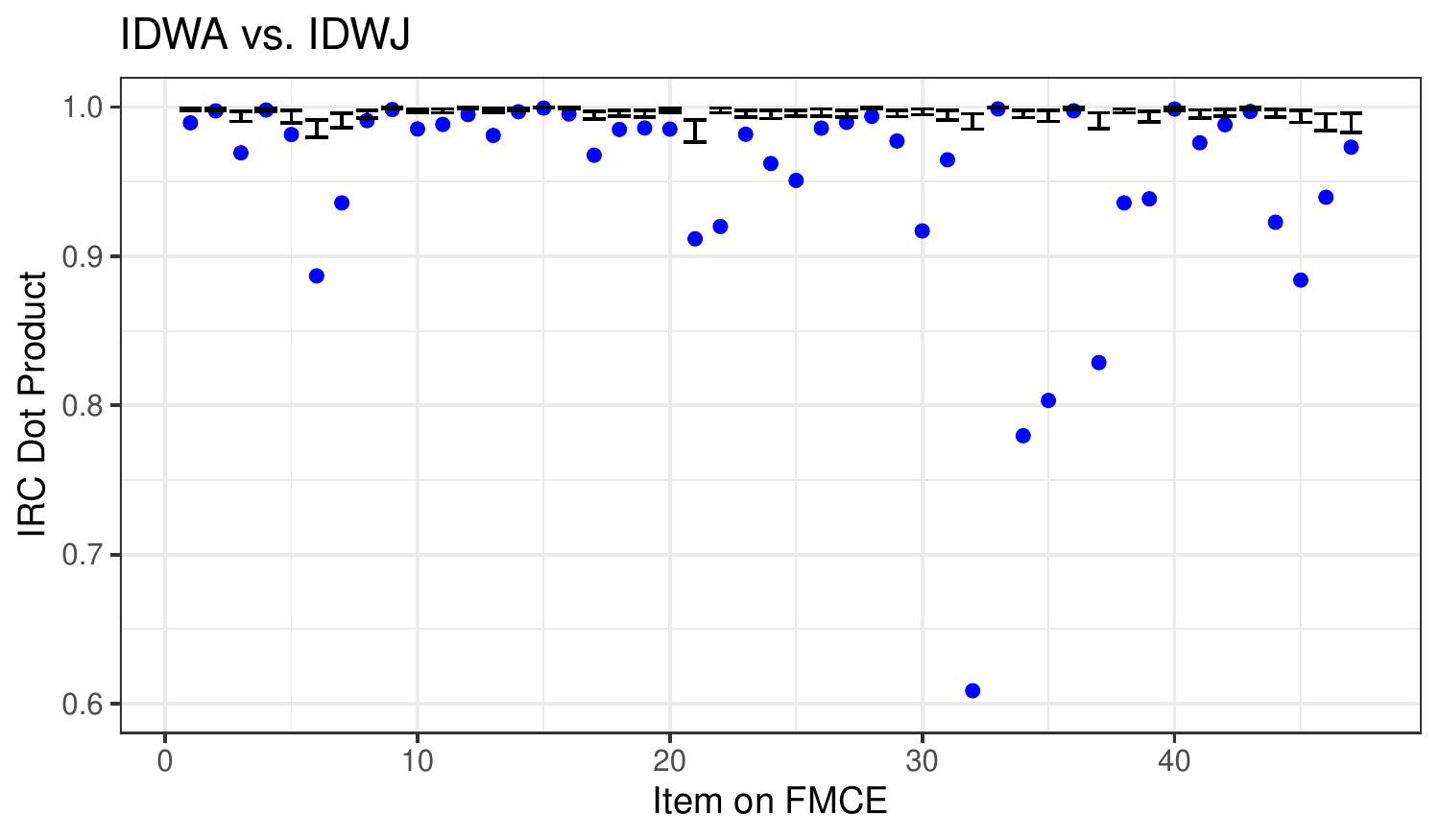}
  \caption{The IRC dot product for each item on the FMCE is shown in blue for the IDW American (IDWA) and Japanese (IDWJ) data sets \cite{Ishimoto2017}. The black error bars represent the randomized trial confidence intervals described in Sec.~\ref{sec:random_interval}.}
  \label{fig:dot_IDWJ_IDWA}
\end{figure}

Figure \ref{fig:dot_IDWJ_IDWA} shows the IRC dot product for each item comparing the Japanese (IDWJ) and American (IDWA) student populations used in the IDW study \cite{Ishimoto2017}. The black error bars in Fig.\ \ref{fig:dot_IDWJ_IDWA} show the randomized trial confidence interval for each item.
The IRC dot product values for 44 of the 47 items fall outside the range of their respective randomized trial confidence intervals, suggesting that the differences between the data sets are likely not due to random chance and represent meaningful differences in student response patterns. The only items with IRC dot product values within their randomized trial confidence interval are items 4, 15, and 40.

As mentioned above, IDW reported 16 items that they noticed having different IRCs between their two data sets: 3, 7, 11--13, 16, 18, 22, 30--32, 34, and 44--47 \cite{Ishimoto2017}. All of these items have IRC dot product values below their randomized trial confidence intervals in Fig.\ \ref{fig:dot_IDWJ_IDWA}, and many of them are well outside the range (e.g., item 34 has an IRC dot product value of 0.780, and a randomized trial confidence interval of [0.993, 0.998]). We find some consistencies when looking at specific items.
\begin{itemize}
    \item Items 3 and 7 have the lowest IRC dot product values of the Force Sled item cluster (items 1--4, 7).
    \item Items 11--13 have lower IRC dot product values than items 8--10.  The second item has the highest value for each of these groups, and the third item has the lowest value.
    \item Item 22 has a lower IRC dot product value than any of the other items in the Acceleration Graphs cluster (items 22--26).
    \item The items involving Newton's third law in the context of cars and trucks (30--38) include those with the lowest IRC dot product values. Items 32 and 34 have the lowest IRC dot product values, indicating the largest differences between the IRCs, and item 31 has a much higher value, indicating better agreement.
\end{itemize}
IDW report that items 16 and 18 may be easier for American students than Japanese students.  Both of these Force Graphs cluster items have IRC dot product values below their respective randomized trial confidence intervals.  Item 21 has the lowest IRC dot product value in the Force Graphs cluster (items 14, 16--21) and thus the most pronounced differences in its IRCs in the region where most students scored.  Visual inspection of the item 21 IDWJ-IDWA IRCs, which are highly similar to the item 21 RSW-IDWJ IRCs shown in Fig.\ \ref{fig:ircs_q21_rsw_idwj}, show that item 21 is easier for both the RSW and IDW American students than the IDW Japanese students.

Figure \ref{fig:dot_IDWJ_IDWA} also reveals additional items that seem to have significantly different IRCs between the two data sets. For example, item 38 has a fairly low IRC dot product of 0.936, which is well outside the random trial confidence interval [0.996, 0.999], but item 38 is not highlighted in the IDW study as being particularly different between American and Japanese students. In fact, IDW mention items 36 and 38 together as showing similarities between their two populations; our IRC dot product analyses support this conclusion for item 36 (IRC dot product of 0.997), but not for item 38. This suggests that some differences between IRCs may be less salient upon visual inspection.

\subsubsection{Comparing two American data sets}
\begin{figure*}[tb]
\begin{minipage}{.45\textwidth}
  \includegraphics[width=\linewidth]{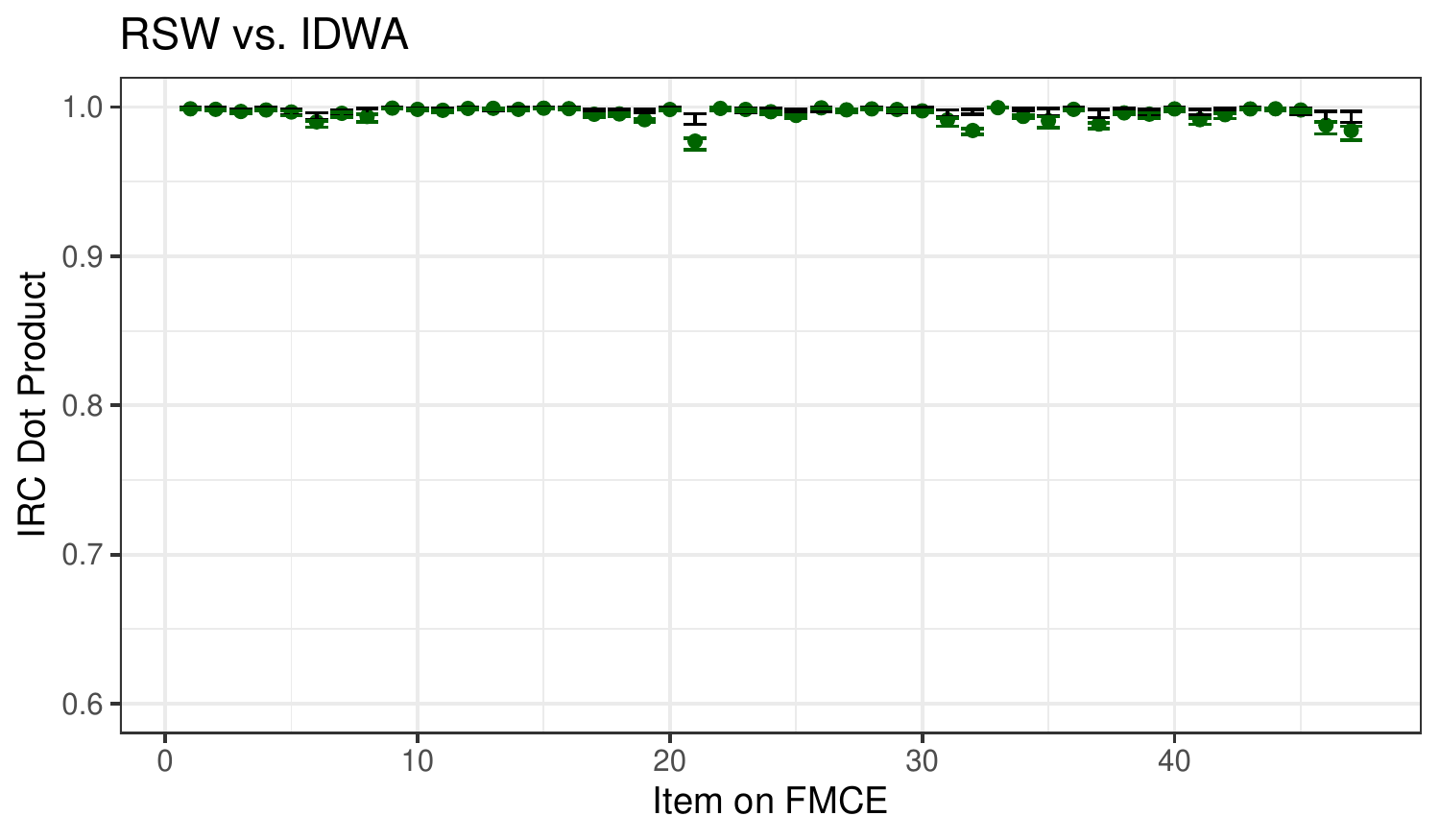}
  
    ~~(a) 
 \end{minipage}
~~~~
\begin{minipage}{.45\textwidth}
  \includegraphics[width=\linewidth]{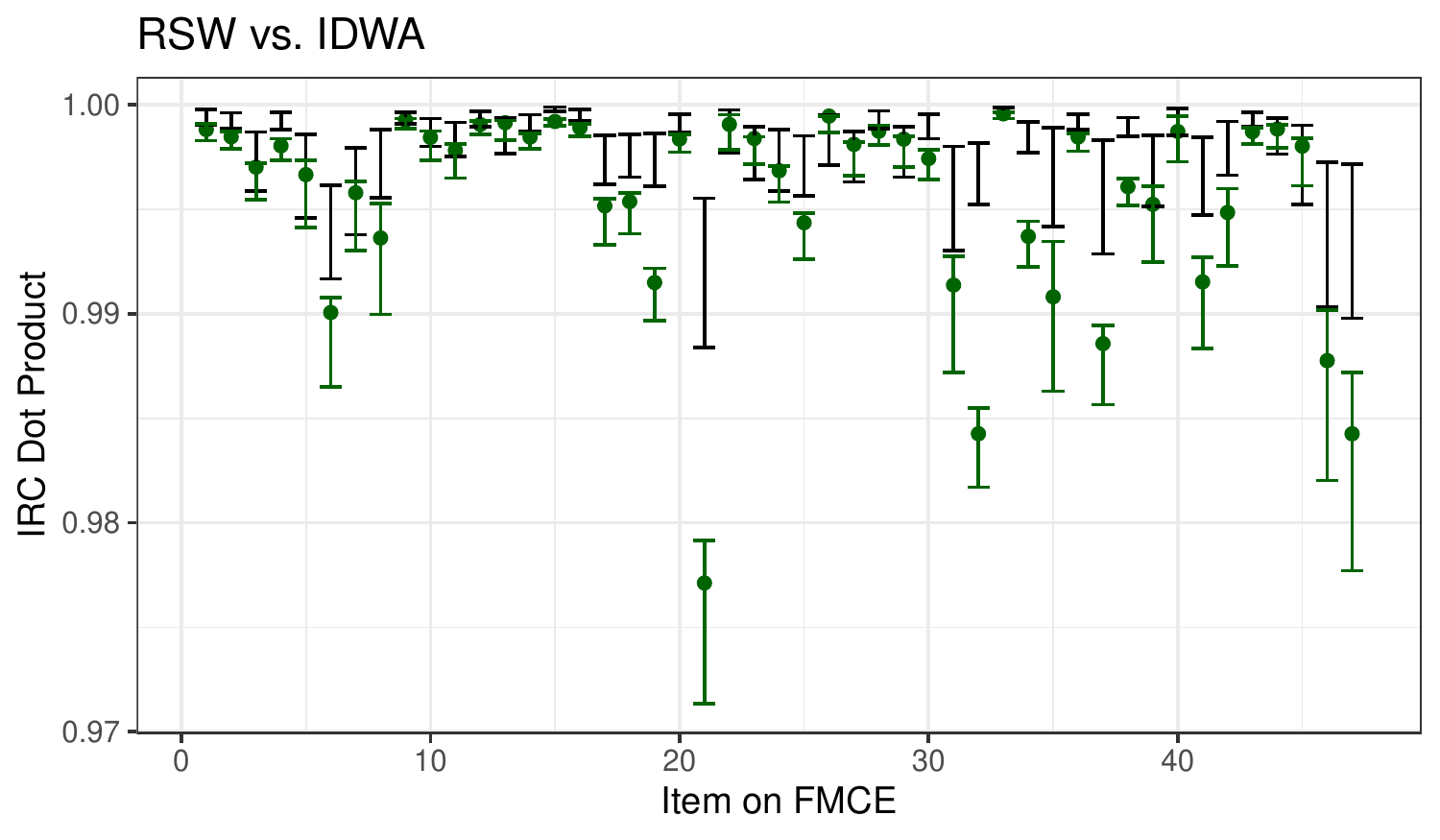}
  
  ~~~~(b) 
 \end{minipage}
\caption{The IRC dot product for each item on the FMCE is shown in green for our American data set (RSW) and the IDW American data set (IDWA). The green error bars around the data points represent the IRC dot product confidence intervals.  The black error bars represent the randomized trial confidence intervals described in Sec.~\ref{sec:random_interval}. The vertical range of (a) matches Figs.\ \ref{fig:dot_IDWJ_IDWA} and \ref{fig:dot_RSW_IDWJ}. Plot (b) contains the same information as (a), but a smaller vertical range is chosen to allow better visualization.}
\label{fig:dot_RSW_IDWA}
\end{figure*}

Figure \ref{fig:dot_RSW_IDWA} shows the IRC dot products comparing the IRCs of both of the American data sets: our RSW data set and the IDWA data set from Ref.\ \cite{Ishimoto2017}. The vertical range of Fig.\ \ref{fig:dot_RSW_IDWA}a was selected to allow easy comparisons with Figs.\ \ref{fig:dot_IDWJ_IDWA} and \ref{fig:dot_RSW_IDWJ}, even though the minimum IRC dot product value is 0.977 (item 21). Figure \ref{fig:dot_RSW_IDWA}b shows the same information as Fig.\ \ref{fig:dot_RSW_IDWA}a, but with a smaller vertical range to allow a more detailed examination of the results. The IRC dot product values of 24 of the 47 items fall within their randomized trial confidence intervals, indicating that any differences between them are potentially the result of random chance.  The IRC dot product values of 22 items are below their randomized trial confidence intervals.  One item (item 26) has an IRC dot product value above its randomized trial confidence interval, indicating a high similarity between those IRCs.

To quantitatively compare the IRCs, we define a comparison metric that is similar to an effect size. The ``dot product effect size'' (DES) for item $i$ is defined as,
\begin{equation}
    DES_i \equiv \frac{RT_{mid,\, i} - \langle\mathbf{a}\cdot\mathbf{b}\rangle_i}{CI_{pooled,\, i}},
\end{equation}
where $RT_{mid,\, i}$ is the midpoint of the randomized trial confidence interval for item $i$, and $\langle\mathbf{a}\cdot\mathbf{b}\rangle_i$ is the IRC dot product between populations $a$ and $b$ for item $i$. The pooled confidence interval $CI_{pooled,\, i}$ is defined as,
\begin{equation}
    CI_{pooled,\, i} \equiv \sqrt{CI_{RT,\, i}^2 + CI_{dot,\, i}^2},
\end{equation}
where $CI_{RT}$ is the size of the randomized trial confidence interval (e.g., the black error bars in Fig.\ \ref{fig:dot_RSW_IDWA}) and $CI_{dot}$ is the size of the IRC dot product confidence interval (e.g., the green error bars in Fig.\ \ref{fig:dot_RSW_IDWA}) \footnote{Our ``dot product effect size'' is not an effect size by Cohen's traditional definition because he uses standard deviation as the scaling factor in the denominator \cite{Cohen1988}. To get a traditional effect size, one must multiply the DES by a factor of 3.92.}. 

\begin{table}[tb]
    \centering
    \begin{tabular}{c@{~~}r@{.}lr@{.}l@{\hspace{1.5cm}}c@{~~}r@{.}lr@{.}l}
    \hline\hline
item&\multicolumn{2}{p{1.25cm}}{RSW vs.\ IDWA}&\multicolumn{2}{p{1.25cm}}{RSW vs.\ IDWJ}&item&\multicolumn{2}{p{1.25cm}}{RSW vs.\ IDWA}&\multicolumn{2}{p{1.25cm}}{RSW vs.\ IDWJ}\\
\hline
1&0&6&4&2&  25&0&8&3&2\\
2&0&7&1&0&  26&-0&5&2&1\\
3&0&1&\textbf{3}&\textbf{0}&  27&-0&2&1&1\\
4&0&9&1&4&  28&0&4&3&0\\
5&0&0&1&9&  29&-0&2&2&7\\
6&0&6&6&6&  30&0&8&\textbf{7}&\textbf{2}\\
7&0&0&\textbf{4}&\textbf{2}&  31&0&6&\textbf{2}&\textbf{8}\\
8&0&6&1&4&  32&2&6&\textbf{13}&\textbf{1}\\
9&0&1&1&1&  33&0&5&1&2\\
10&0&1&4&1& 34&1&8&\textbf{12}&\textbf{7}\\
11&0&2&\textbf{2}&\textbf{3}& 35&0&7&4&7\\
12&0&3&\textbf{2}&\textbf{7}& 36&0&7&1&7\\
13&-0&3&\textbf{2}&\textbf{7}& 37&1&1&4&7\\
14&0&6&0&8& 38&1&8&10&9\\
15&1&5&0&0& 39&0&3&3&3\\
16&0&8&\textbf{0}&\textbf{7}& 40&0&2&0&4\\
17&0&7&3&5& 41&0&9&2&3\\
18&0&8&\textbf{0}&\textbf{9}& 42&0&7&0&7\\
19&1&6&1&4& 43&0&5&1&0\\
20&0&6&6&2& 44&-0&2&\textbf{6}&\textbf{7}\\
21&1&4&3&9& 45&-0&2&\textbf{5}&\textbf{4}\\
22&-0&1&\textbf{5}&\textbf{5}& 46&0&6&\textbf{3}&\textbf{4}\\
23&-0&2&1&8& 47&0&8&\textbf{1}&\textbf{9}\\
24&0&1&2&7&&&&&\\
\hline
    \end{tabular}
    \caption{The dot product effect size (DES) values for all comparisons between our RSW data set, and the IDW American (IDWA) and Japanese (IDWJ) data sets. Bolded entries indicate items that were highlighted by IDW as being particularly different between their American and Japanese data sets.}
    \label{tab:dotEffect}
\end{table}

Table \ref{tab:dotEffect} shows the DES value for each item for the comparison between our RSW data set and the IDW American data set (RSW vs.\ IDWA columns). Comparing Table \ref{tab:dotEffect} with Fig.\ \ref{fig:dot_RSW_IDWA} provides a visual interpretation for our DES values: items 8, 31, and 46 all have error bars that are just barely touching, and they all have DES values of 0.6 \footnote{When the confidence intervals are just barely touching, the DES value would be 1 if the IRC dot product value was at the center of the IRC dot product confidence interval.  When the IRC dot product value is close to 1, it will be higher than the midpoint of the IRC dot product confidence interval due to ceiling effects.}. Over half of the items (28 out of 47) have DES values at or below 0.6, indicating that the error bars overlap, and there is a reasonable likelihood that the value of the IRC dot product is high enough for the RSW and IDWA IRCs to be considered the same. Of the remaining 19 items, only one (item 32) has a DES value above 2. Additionally, nine items have negative DES values, indicating that the IRC dot products for those items are higher than the midpoint of the respective randomized trial confidence intervals, suggesting that those IRCs are highly similar.

\subsubsection{Comparing a new American data set to the previous Japanese data set}
\begin{figure}[htp!]
  \includegraphics[width=\linewidth]{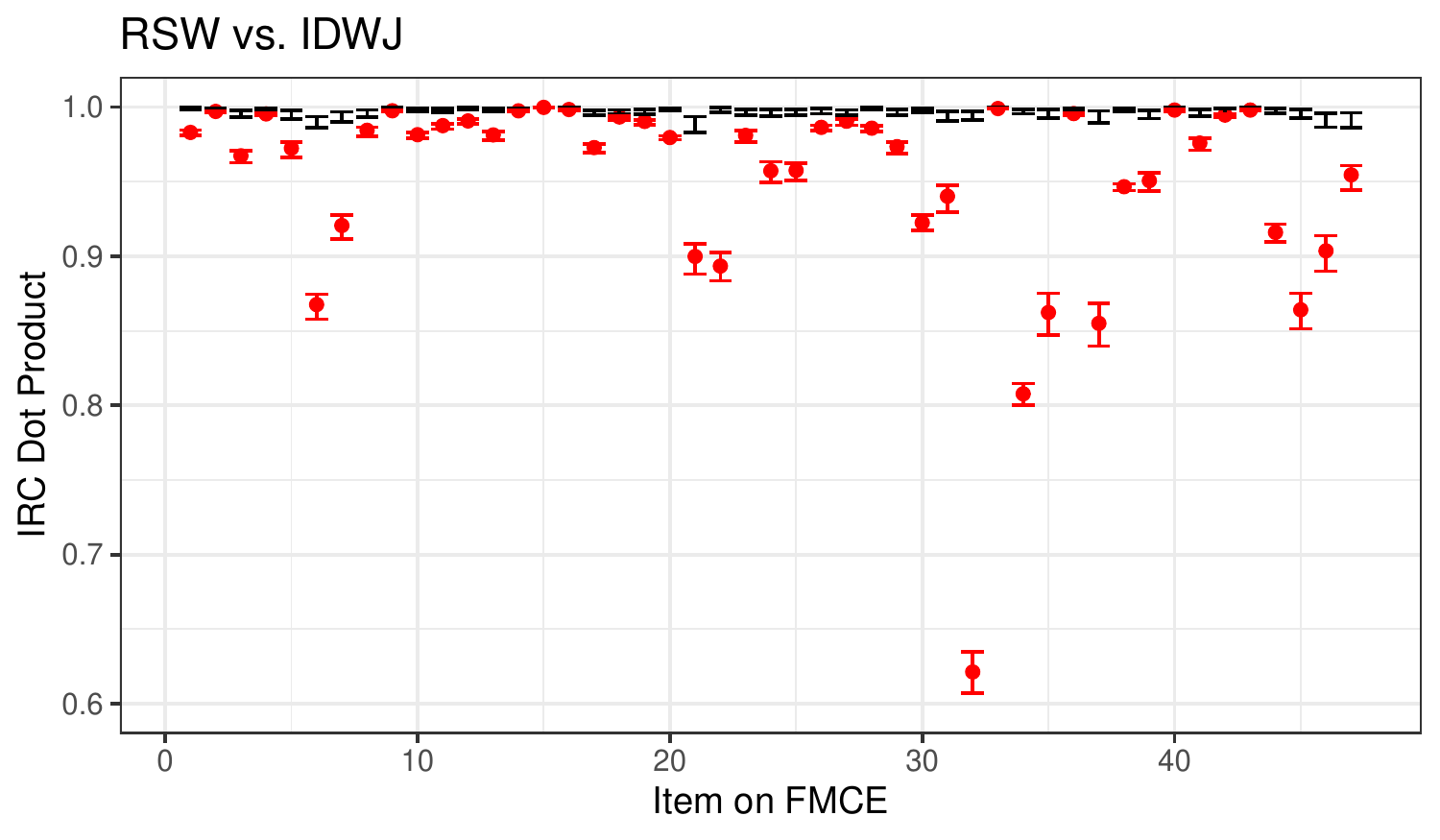}
  \caption{The IRC dot product for each item on the FMCE is shown in red for this our American data set (RSW) and the IDW Japanese data set (IDWJ). The red error bars around the data points represent the IRC dot product confidence intervals. The black error bars represent the randomized trial confidence intervals described in Sec.~\ref{sec:random_interval}.}
  \label{fig:dot_RSW_IDWJ}
\end{figure}

Figure \ref{fig:dot_RSW_IDWJ} shows the IRC dot products comparing the IRCs of our American data set (RSW) and the IDW Japanese data set. These are highly similar to the IRC dot products shown in Fig.\ \ref{fig:dot_IDWJ_IDWA} comparing the IDW Japanese and American students. 
This result is to be expected given that the IRC dot products shown in Figs.\ \ref{fig:dot_all} and \ref{fig:dot_RSW_IDWA} show the IRCs of both American populations are highly similar. For the data sets in Fig.\ \ref{fig:dot_RSW_IDWJ}, 46 out of the 47 items have IRC dot product values below the range of their respective randomized trial confidence intervals (only item 15 is within the confidence interval). 

\begin{figure}
    \centering
    \includegraphics[width = \columnwidth]{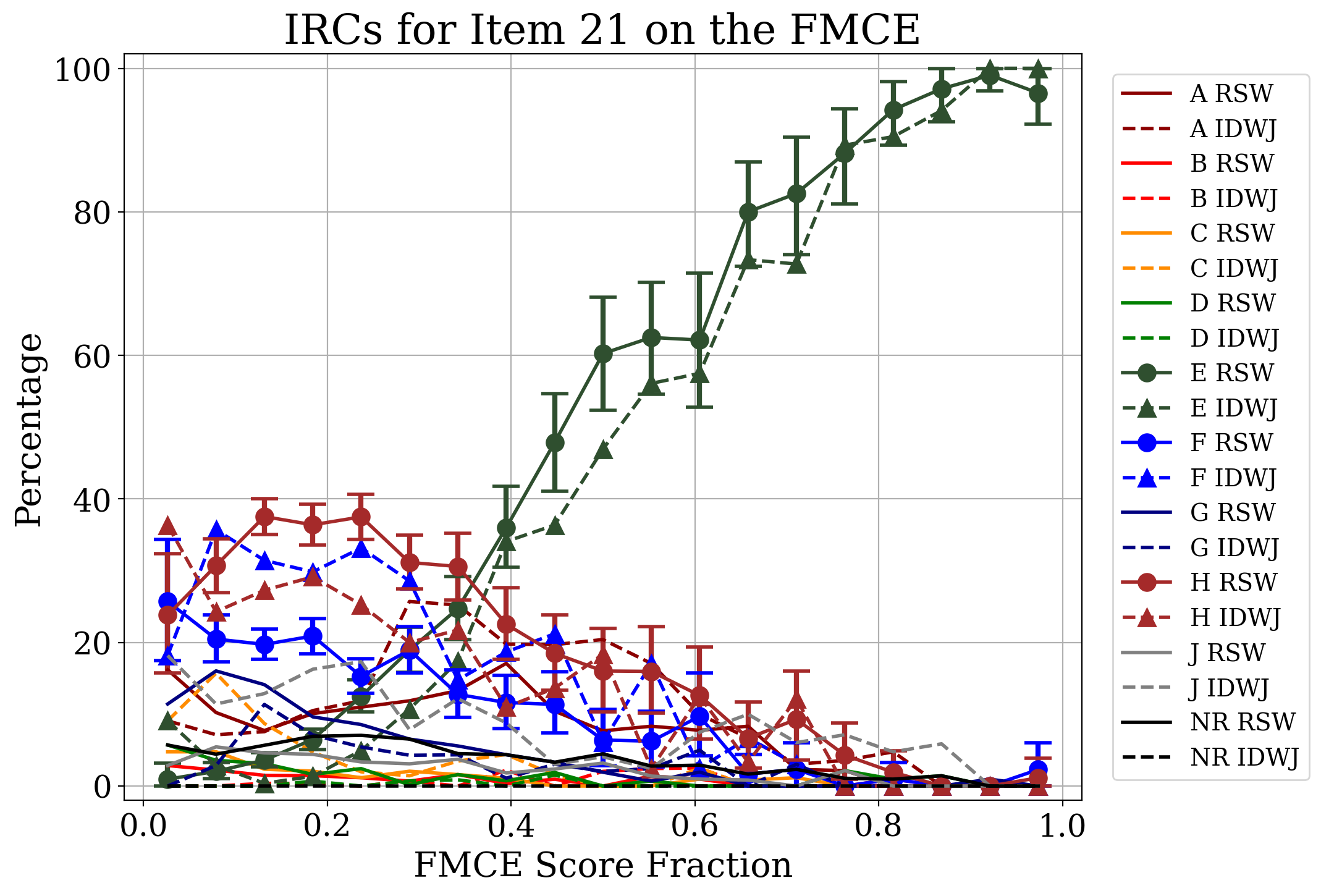}
    \caption{Item 21 IRCs: the solid lines show the IRCs for the RSW data set, and the dashed lines show the IRCs for the IDWJ data set. NR = no response.  The error bars on E, F, and H RSW IRCs represent the IRC dot product confidence intervals.}
    \label{fig:ircs_q21_rsw_idwj}
\end{figure}

Table \ref{tab:dotEffect} also shows the DES value for each item for the comparison between the RSW and IDW Japanese (IDWJ) data sets. The DES values in the RSW vs.\ IDWJ columns are generally larger than the DES values in the RSW vs.\ IDWA columns. Only three items have smaller DES values, and two of these (items 16 and 19) are within 0.2 of the RSW vs.\ IDWA value \footnote{Item 15 seems to be an outlier in that the DES for the IDWJ comparison is much smaller than the IDWA comparison. We believe this to be attributable to a ceiling effect: the randomized trial confidence interval for item 15 in Fig.\ \ref{fig:dot_RSW_IDWJ} has a range of [0.9996, 0.9999]. Item 15 is one of the easiest on the FMCE, with most students answering the item correctly before instruction, which is why item 15 is typically omitted from calculations of an overall score.}. The median DES value is 2.7 for the RSW/IDWJ comparison (compared to 0.6 for RSW/IDWA), and the highest value is 13.1 (again for item 32). Over half of the items (29 out of 47) have DES values above 2, compared to only one item for the RSW vs.\ IDWA results. Only two items (15 and 40) have DES values less than 0.6 (interpreted above as having a reasonable expectation of having similar IRCs) \footnote{Items 15 and 40 are two of the easiest items on the FMCE, with most students answering them correctly before instruction: 94\% correct on item 15 and 86\% correct on item 40 in our RSW data set.}.

The items highlighted as being notably different by IDW (mentioned above) are shown in bold in Table \ref{tab:dotEffect}. These include the two highest DES values (items 32 and 34), but they do not include all of the highest values (e.g., item 38 is omitted from their discussion). Quantitatively comparing IRCs by computing the IRC dot product, combined with our methods for generating confidence intervals, can reveal different patterns than qualitative visual comparisons. Of particular interest are items 16 and 18, which IDW highlighted as seeming easier for American students. Similar to what we mentioned above with the IDWA vs.\ IDWJ comparison, we find other items in the Force Graphs cluster (items 14, 16--21) to have greater differences than items 16 and 18; moreover, Table \ref{tab:dotEffect} also shows that the RSW vs.\ IDWJ comparisons for items 16 and 18 are very similar to the RSW vs.\ IDWA comparisons. The differences reported by IDW for these two items may not be generalizable for all American students.  Visual inspection of Fig.\ \ref{fig:ircs_q21_rsw_idwj} shows that item 21 seems to be somewhat easier for the RSW American students than the IDW Japanese students: the RSW IRC for the correct response E is slightly to the left of the IDWJ IRC, and more RSW students select E in the most populated region of the plot (fractional scores between about 0.1 and 0.3). The absence of a single dominant incorrect response and the differences in selected answer choices in the populated region lead to a higher DES value than either item 16 or 18 (see Table \ref{tab:dotEffect}).

\subsection{Examining specific items}
\begin{figure}[tb]
  \includegraphics[width=\linewidth]{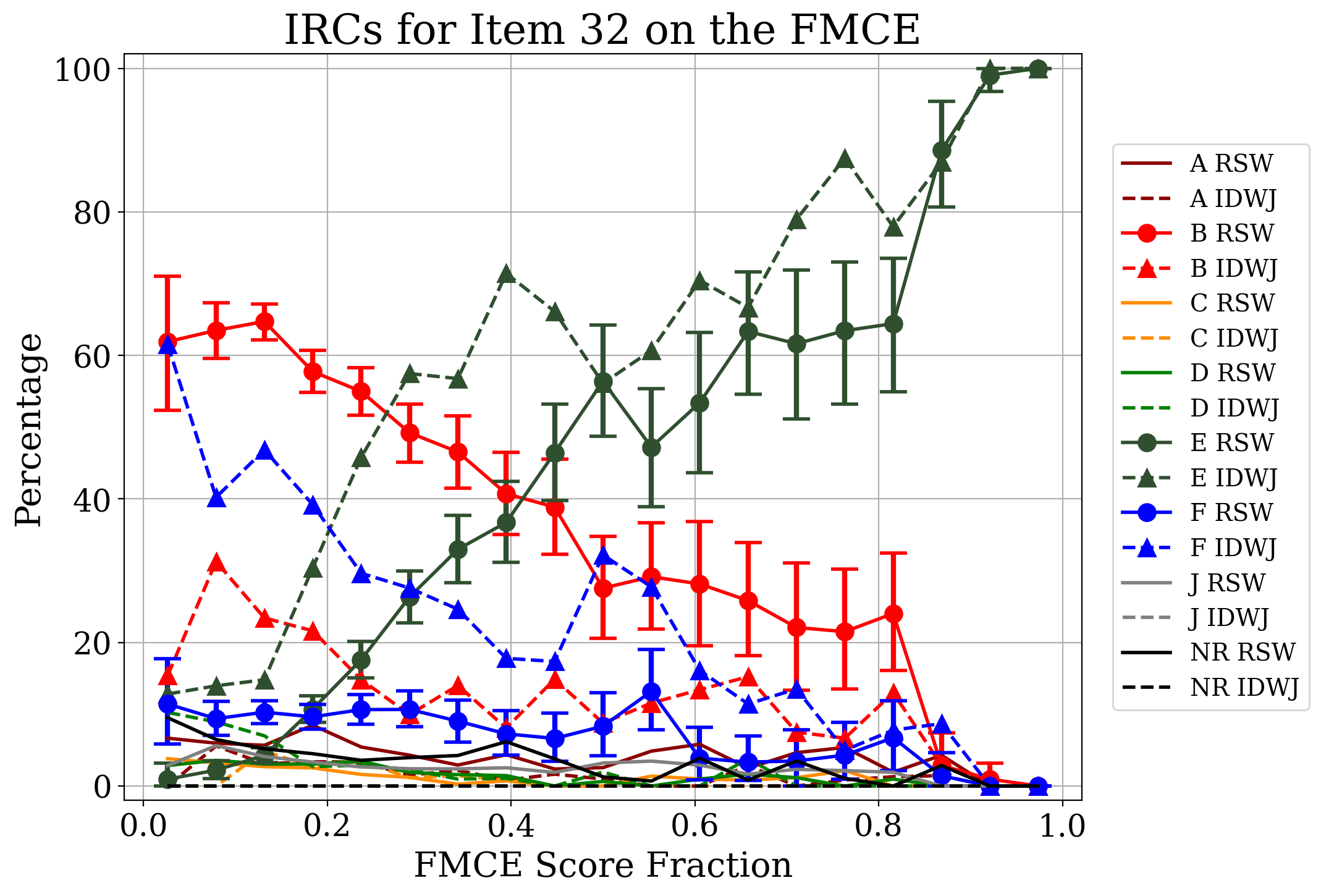}
  \caption{Item 32 IRCs: the solid lines show the IRCs for the RSW data set, and the dashed lines show the IRCs for the IDWJ data set. NR = no response.  The error bars on B, E, and F RSW IRCs represent the IRC dot product confidence intervals.}
  \label{fig:ircs_q32_rsw_idwj}
\end{figure}

The IRC dot product for item 32 between an American data set and the Japanese data set is the lowest value of any item on the FMCE (0.621 for RSW vs.\ IDWJ and 0.609 for IDWA vs.\ IDWJ), suggesting the IRCs differ more for this item than any other. Figure \ref{fig:ircs_q32_rsw_idwj} shows us that the American students in the RSW data set are more likely to choose B (consistent with the IDWA results) while the Japanese students (IDWJ) are more likely to choose F.  Item 32 involves a small car colliding with a large stationary truck. Answer B indicates that the car exerts a larger force on the truck than the truck on the car, which is consistent with the idea that faster, or more active, objects exert more force (a.k.a.\ action dependence \cite{Smith2008}).  Answer F indicates that there is not enough information to determine the relative magnitudes of the interaction forces between the car and the truck. IDW claim that this is most likely due to different life experiences between American and Japanese students, where Americans are more likely to grow up using automobiles for commuting. 
Our results confirm that large portions of American students choose answer B. The most considerable differences between Japanese and American students' responses occur at low scores, which contain most of the students (see Fig.\ \ref{fig:score_distributions}).   

\begin{figure}[bt]
  \includegraphics[width=\linewidth]{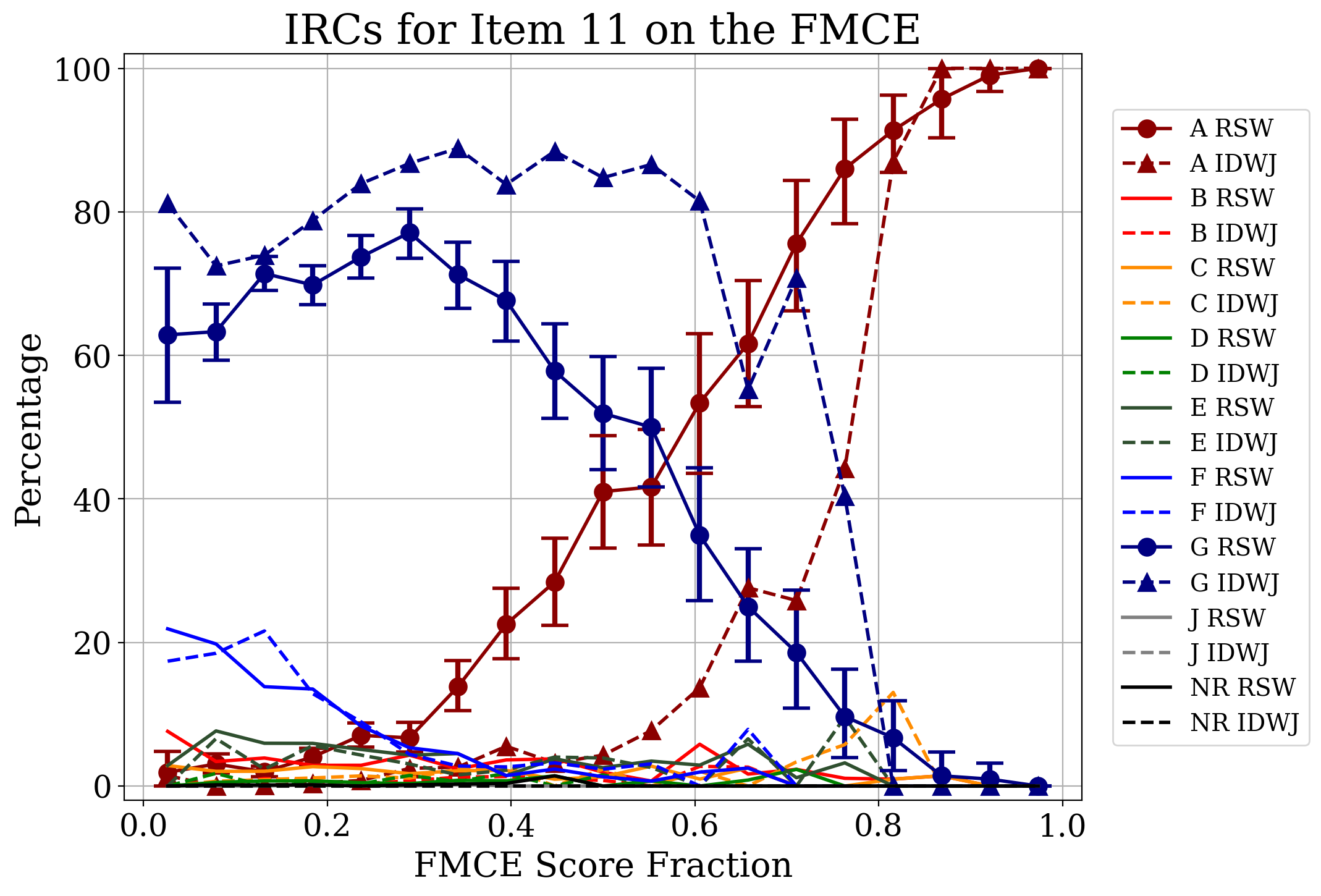}
  \caption{Item 11 IRCs: the solid lines show the IRCs for the RSW data set, and the dashed lines show the IRCs for the IDWJ data set. NR = no response.  The error bars on A and G RSW IRCs represent the IRC dot product confidence intervals.}
  \label{fig:ircs_q11_rsw_idwj}
\end{figure}

Figure \ref{fig:ircs_q11_rsw_idwj} shows the IRCs of our American (RSW) and the IDW Japanese (IDWJ) data sets for item 11, which examines the force acting on a coin tossed upward while still ascending.  From visual inspection, we see that the IRCs that differ the most are for answer choices A (the correct answer choice) and G (the most common incorrect answer) over the fractional score range of 0.3 -- 0.8; however, the IRC dot product for this pair of populations on item 11 is 0.988, suggesting that the IRCs are much more similar than for item 32.  While the large differences in IRC dot product values between item 11 and 32 may not be evident from viewing the IRCs, the IRC dot product value depends on each population's score distributions. The score distributions' peaks are below 0.2 for all three data sets, with over half of the students in the fractional score range 0.1 -- 0.3. The large differences between the item 32 IRCs in this highly-populated range lead to a much lower IRC dot product value.

Similar comparisons can be made between the IRCs for item 11 (Fig.\ \ref{fig:ircs_q11_rsw_idwj}) and item 21 (Fig.\ \ref{fig:ircs_q21_rsw_idwj}). It would be reasonable to look at these two plots and conclude that the two populations are more different on item 11 than they are for item 21: the difference between the RSW and IDWJ IRCs for the dominant (and correct) answer choice for item 21 (E) appear much smaller than the differences between the two dominant answer choices for item 11; however, Table \ref{tab:dotEffect} shows that item 21 has a higher DES value, and is, therefore, less likely to be attributed to random chance. To resolve this apparent discrepancy, we must again focus on the plot's most populous region: scores from 0.1--0.3. In this region item 11 shows three IRCs with non-negligible percentage of being chosen: correct (A), dominant incorrect (G), and one other incorrect (F), which is mostly chosen by the lowest-scoring students. In contrast, item 21 seems to have five IRCs showing answers chosen by more than 10\% of students in each data set in this score range, several with notable differences between the data sets. In the highly-populated score range of 0.1--0.3, we see the biggest differences between RSW and IDWJ on item 32, followed by item 21, and the most similarity (of these three items) with item 11; this is consistent with both the IRC dot product values in Fig.\ \ref{fig:dot_RSW_IDWJ} (0.621, 0.900, and 0.988 for items 32, 21, and 11, respectively) and the DES results in Table \ref{tab:dotEffect} (13.1, 3.9, and 2.3 for items 32, 21, and 11, respectively).

\section{Conclusions}
\label{sec:conclusion}
Our analyses support the results reported by Ishimoto, Davenport, and Wittmann \cite{Ishimoto2017}: American physics students' and Japanese students' preinstruction responses to the Force and Motion Conceptual Evaluation differ for many items in non-trivial ways. The majority of these differences are related to specific choices of incorrect answers rather than differences in numbers of correct answers. One limitation of our study is that we could not fully replicate IDW's work using a disjoint data set of Japanese physics students' responses.  However, our work strongly suggests that the IDW American data set is representative of American students in general. One caveat to this claim is the well-documented tendency for physics education research studies to oversample from large research universities, which tend to have students with more previous exposure to formal physics instruction than is typical for American college and university physics students in general \cite{Kanim2020}.  Unfortunately, due to the anonymous nature of the majority of our data set (provided by the PhysPort Data Explorer), we are unable to determine whether or not our student population is truly representative of all American physics students; however, the similarity between our results and the IDW study are quite suggestive.

In terms of score distributions, our data set fell between the IDW American and Japanese data sets. Similar to IDW, we find these differences to be statistically significant ($p<0.001$ for both the Kruskal-Wallace, and the pairwise Wilcoxon rank-sum tests), but we do not consider these differences to be pedagogically meaningful. The differences in the mean and median scores between all three data sets are within about 2 out of 37 points, and the effect sizes are small. Additionally, looking at these score distributions alone would likely not impact an instructor's approach to teaching without having more context highlighting the differences between the populations.

The pedagogically important differences between student populations can be seen in the IRCs. For the vast majority of items, the RSW IRCs are very similar to the IRCs for the IDW American data set. 
IRC dot product values are very high overall, and random chance can account for the differences between IRCs for many of the items. 
Conversely, the IRC dot product values between the RSW and the IDW Japanese data sets are much lower for many items, and the dot product effect size (DES) values suggest that these differences are not the result of random fluctuations but are likely due to actual differences in the ways that the two student populations select answer choices for specific FMCE items. 
Additionally, the IRC dot product values comparing the RSW and IDW Japanese data sets are very similar to the IRC dot product values comparing the IDW American and Japanese data sets, providing additional support to our claim that our results largely support those reported by IDW.

Of items in the Force Graphs Cluster, IDW reported that items 16 and 18 may be slightly easier for American students based on visual qualitative comparisons of the IRCs \cite{Ishimoto2017}. We find that item 21 is slightly easier for American students more so than items 16 and 18. IDW reported that items 16 and 18 may be slightly easier for American students based their claims on visual qualitative comparisons of the IRCs \cite{Ishimoto2017}. When visually comparing IRCs, we focus primarily on the region where majority of students scored---the vast majority earn fractional scores less than 0.4 (15 out of 37). Differences between IRCs are more meaningful when they are representative of more students, and a benefit of computing the IRC dot product is that each score bin is weighted by the number of students. Using qualitative visual comparisons and quantitative IRC dot products can provide a robust way to compare IRCs and gain valuable insights into similarities and differences between how different groups select answers to the FMCE and other multiple-choice test items.

\section{Implications for Instruction}
Our data set from many American colleges and universities is similar to the disjoint IDW data set from many American colleges and universities.  This is not to say that every class of physics students at an American college or university will select the same proportions of specific answers to the FMCE as shown in our data. One of the strongest results we see from comparing Japanese and American students is that students with different backgrounds answer the FMCE differently, and these differences do not necessarily show up in score distributions that focus only on the number of items answered correctly.

As mentioned above, slight differences in preinstruction score distributions may have little impact on instructional methods, but large differences in the incorrect responses students select could (and should) inform an instructor's approach to a class. Figure \ref{fig:ircs_q32_rsw_idwj} illustrates this wonderfully: the RSW American students mostly choose one incorrect answer for item 32 (B, consistent with the idea that faster objects exert more force than slower objects), but the IDW Japanese students mainly select a different choice (F, indicating that more information is needed to determine the forces). Both of these answers are incorrect, but the difference could represent vastly different ways of thinking about the forces involved in two-body collisions; moreover, the differences between these incorrect IRCs are more pronounced than the correct IRCs (answer E), suggesting that an inspection of correct answers alone may hide the magnitude of the differences between these populations. 
An essential feature of good physics instruction is acknowledging students' initial ideas and using them to facilitate growth in understanding. Looking only at correct answers cannot accomplish this, but IRCs can show if different groups of students choose different answers (e.g., those with low, middle, and high scores).

\begin{acknowledgments}
	We thank Sam McKagan and Ellie Sayre for providing access to data from PhysPort's Data Explorer and all of the instructors who were willing to share their students' FMCE responses. We also thank Nicholas Baltera, Paul Kelly, Maria Lentini, and Mitchell Nussenbaum for their previous contributions as research team members. We are deeply grateful to Glen Davenport for his assistance on this project. The National Science Foundation supported this project through grant DUE-1836470. 
\end{acknowledgments}

\bibliography{references.bib}

\end{document}